\documentclass[a4paper,11pt]{article}
\usepackage{graphicx}
\usepackage{hyperref}
\title{Retrofitting mutual authentication to GSM using RAND hijacking}
\author{Mohammed Shafiul Alam Khan\thanks{The first author is a commonwealth scholar, funded by the UK Government}~~and Chris J. Mitchell\\
Information Security Group, Royal Holloway, University of London\\
{\tt shafiulalam@gmail.com}; {\tt me@chrismitchell.net}}

\begin{document}

\bibliographystyle{plain}

\maketitle

\section*{Abstract}

As has been widely discussed, the GSM mobile telephony system
only offers unilateral authentication of the mobile phone to
the network; this limitation permits a range of attacks.  While
adding support for mutual authentication would be highly
beneficial, changing the way GSM serving networks operate is
not practical.  This paper proposes a novel modification to the
relationship between a Subscriber Identity Module (SIM) and its
home network which allows mutual authentication without
changing any of the existing mobile infrastructure, including
the phones; the only necessary changes are to the
authentication centres and the SIMs. This enhancement, which
could be deployed piecemeal in a completely transparent way,
not only addresses a number of serious vulnerabilities in GSM
but is also the first proposal for enhancing GSM authentication
that possesses such transparency properties.

\section{Introduction} \label{Intro}

This paper proposes a way of adding network-to-phone
authentication to the GSM mobile phone system, in a way that is
completely transparent to the existing network infrastructure.
Currently, GSM only supports authentication of the phone to the
network, leaving the system open to a wide range of threats
(see, for example, \cite{K353a}).  Despite the introduction and
deployment of 3G (UMTS) and 4G (LTE) mobile phone systems,
which rectify the GSM problem by providing mutual
authentication between phone and network, GSM remains of huge
practical importance worldwide and is not likely to be replaced
for many decades to come.  As a result, finding ways of
improving the security offered by GSM, without the need for
changes to the deployed phones and access networks, is clearly
of great practical significance. This observation motivates the
work described in this paper.

It is somewhat counterintuitive to propose that authentication
of the network to the phone can be achieved without modifying
the way in which the existing network and phones operate.  This
apparently paradoxical result is achieved by using a technique
we refer to as \emph{RAND hijacking}.  This involves using the
\emph{RAND} value, which serves as a nonce in the existing
unilateral authentication protocol and is sent from the network
to the phone, to contain data which enables the recipient SIM
to verify its origin and freshness.  That is, the \emph{RAND}
is hijacked to act as a communications channel between a home
network and a SIM.

The remainder of the paper is structured as follows.  Key facts
about the GSM network, including details of the operation of
the GSM authentication and key establishment (AKA) protocol,
are given in section~\ref{GSM}. This is followed in
section~\ref{hijacking} by an introduction to the notion of
RAND hijacking.  In section~\ref{auth}, the novel enhanced
version of the GSM authentication scheme is described, and
section \ref{using} describes how the SIM can use the results
of the network authentication to affect UE behaviour.  An
analysis of the novel system is provided in
section~\ref{analysis}.  The relationship of the proposed
scheme to the prior art is discussed in section~\ref{prior},
and the paper concludes in section~\ref{conclusion}.

\section{GSM}  \label{GSM}

\subsection{Terminology}

We start by providing a brief overview of key terminology for
mobile systems.  We focus in particular on the GSM network, but
much of the description applies in slightly modified form to 3G
and 4G networks.  A more detailed description of GSM security
features can, for example, be found in Pagliusi
\cite{Pagliusi02}.

A complete mobile phone is referred to as a \emph{user
equipment (UE)}, where the term encapsulates not only the
\emph{mobile equipment (ME)\@}, i.e.\ the phone, but also the
\emph{subscriber identity module (SIM)} within it, where the
SIM takes the form of a cut-down smart card. The SIM embodies
the relationship between the human user and the issuing
\emph{home network}, including the \emph{International Mobile
Subscriber Identity (IMSI)}\@, the telephone number of the UE,
and other user (subscriber) data, together with a secret key
shared with the issuing network which forms the basis for all
the air interface security features.

To attach to a mobile network, a UE connects via its radio
interface to a radio tower. Several radio towers are controlled
by a single \emph{radio network controller (RNC)} which is
connected to one \emph{mobile switching center/visitor location
register (MSC/VLR)}\@. The MSC/VLR is responsible for
controlling call setup and routing. Each MSC/VLR is also
connected to the carrier network's \emph{home location register
(HLR)} where corresponding subscriber details can be found. The
HLR is associated with an \emph{authentication center (AuC)}
that stores cryptographic credentials required for
communicating with the SIM; specifically, the AuC shares a
unique secret key $K_i$ with each SIM issued by the network to
which it belongs. The RNC and the MSC/VLR are part of the
\emph{visiting/serving network} whereas the HLR and the AuC are
the \emph{home network} component.

\subsection{GSM authentication protocol}  \label{AKA}

To prevent unauthorised mobile devices gaining access to
network service, GSM incorporates an authentication procedure
which enables the network to verify that the SIM in a UE is
genuine. The authentication procedure operates as follows. Further details can be found in technical specifications GSM 03.20 \cite{GSM_03_20} and GSM 04.08 \cite{GSM_04_08}.

\begin{enumerate}
\item The UE visits a network, and is initially identified
    using its IMSI.
\item The visited network identifies the UE's home network
    from the supplied IMSI, and contacts the home network
    for authentication information.
\item The home network's AuC generates one or more
    \emph{authentication triples} (\emph{RAND},
    \emph{XRES}, $K_c$), and sends them to the visited
    network, where \emph{RAND} is a 128-bit random
    `challenge' value, \emph{XRES} is the 64-bit `expected
    response', and $K_c$ is a 64-bit short-term session key
    to be used to encrypt data sent across the air
    interface between the UE and the network.
\item The visited network sends \emph{RAND} to the UE as an
    authentication challenge.
\item The ME receives the \emph{RAND}, and passes it to the
    SIM.
\item The SIM computes \emph{SRES} =
    A3$_{K_i}$(\emph{RAND}) and $K_c$ =
    A8$_{K_i}$(\emph{RAND}), where A3 and A8 are
    network-specific cryptographic functions; A3 is a MAC
    function and A8 is a key derivation function.  Note
    that precisely the same computation was performed by
    the AuC in step 3 to generate \emph{XRES} and $K_c$.
\item The SIM passes \emph{SRES} and $K_c$ to the ME.
\item The ME keeps the session key $K_c$ for use in data
    encryption, and forwards \emph{SRES} to the serving
    network.
\item The serving network compares \emph{SRES} with
    \emph{XRES}; if they are the same the UE is deemed
    authenticated, and $K_c$ can now be used for traffic
    encryption using any of the standardised algorithms
    (i.e.\ one of A5/1, A5/2 and A5/3), as selected by the
    serving network.
\end{enumerate}

\subsection{Vulnerabilities}  \label{vulns}

The GSM AKA protocol clearly only provides one-way
authentication. As widely documented (see, for example,
\cite{K353a}), this permits a `false' base station to
impersonate a genuine network and interact with a UE\@. This in
turn gives rise to a range of security weaknesses.  We are
particularly interested in attacks of the following types.

\begin{itemize}
\item Because the network always decides whether or not to
    enable encryption, it is possible for a malicious party
    to act as an intermediary between a UE and a genuine
    network, impersonating the network to the UE and using
    a genuine SIM of its own to talk to the network.  All
    traffic sent via the man-in-the-middle is simply
    relayed. The false network does not enable encryption
    on the link to the UE, so the fact that it does not
    know the encryption key does not matter.  If the
    genuine network chooses to enable encryption, then the
    man-in-the-middle can communicate with it successfully
    since it is using its own SIM for this leg of the
    communications. As a result, the man-in-the-middle can
    seamlessly listen to all the voice traffic sent to and
    from the victim UE, at the cost of paying for the call.

\item The fact that the network decides whether or not to
    enable data encryption also enables the well known
    Barkan-Biham-Keller attack, \cite{K395}. This attack is
    designed to recover the encryption key $K_c$, and hence
    enable unlimited interception of phone calls.  The
    attack takes advantage of three key facts: A5/2 is very
    weak, the network decides which algorithm to use, and
    the same key $K_c$ is used with all three encryption
    algorithms. One possible scenario for the attack is as
    follows.

Suppose an eavesdropper intercepts the AKA exchange between
the network and a UE (notably including the \emph{RAND}),
and also some of the subsequent encrypted voice exchanges
involving that UE\@. Suppose also that the UE is
subsequently switched on within the range of a fake network
operated by the attacker.  The fake network inaugurates the
AKA protocol with the UE, and sends the previously
intercepted \emph{RAND}\@, causing the SIM in the UE to
generate the same $K_c$ as was used to encrypt the
intercepted data.  The UE responds with \emph{SRES} (which
the fake network ignores) and the fake network now enables
encryption using A5/2. The UE will now send data to the
network encrypted using A5/2 with the key $K_c$; because of
certain details of the GSM protocol, the plaintext data
will contain predictable redundancy. The fake network now
takes advantage of the weakness of A5/2 to recover $K_c$
from the combination of the ciphertext and known redundancy
in the corresponding plaintext.  The key $K_c$ can now be
used to decrypt all the previously intercepted data, which
may have been encrypted using a strong algorithm such as
A5/3.
\end{itemize}

The lack of mutual authentication has been addressed in 3G and
later networks.  As a result it is tempting to suggest that
trying to fix GSM is no longer of relevance.  However, GSM
continues to be very widely used worldwide and will continue to
be for many years to come; so finding ways of upgrading GSM
post-deployment appears to be worthwhile. However, any such
solution must work with the existing infrastructure, i.e.\ the
existing serving network systems.  We are therefore interested
in a solution which only requires SIMs and the home network to
be upgraded.  Such a solution can be rolled out piecemeal with
no impact on the existing global infrastructure, and this is
the focus of the remainder of this paper.

\subsection{Proactive SIM}  \label{proactive}

Before proceeding we need to briefly review a key piece of GSM
technology which enables the SIM to send an instruction to the
ME\@. \emph{Proactive SIM} is a service
operating across the SIM-ME interface that provides a
mechanism for a SIM to initiate an action to be taken by the
ME\@. It forms part of the \emph{SIM application toolkit (STK)}, which was introduced in the GSM 11.14 technical specification \cite{GSM_11_14}. Communications between an ME and a SIM are command/response based, and STK provides a set of commands which allow the SIM to interact and operate with any ME which supports them.

GSM technical specification \cite{GSM_11_11} states that the ME
must communicate with the SIM using either the T=0 or T=1
protocol, specified in ISO/IEC 7816-3 \cite{ISO_IEC_7816-3}. In
both cases the ME is always the \emph{master} and thus
initiates commands to the SIM; as a result there is no
mechanism for the SIM to initiate communications with the
ME\@. This limits the possibility of introducing new SIM
features requiring the support of the ME, as the ME needs to
know in advance what actions it should take. The proactive SIM
service provides a mechanism that allows the SIM to indicate
to the ME, using a response to an ME-issued command, that it has
some information to send. The SIM achieves this by including a
special status byte (`91' followed by the length of the instruction to send) in the response application protocol data
unit. The ME is then required to issue the \emph{FETCH} command
to find out what the information is \cite{ETSI_TS_102_223}. The ME must now execute the SIM-initiated command and return the result in the \emph{TERMINAL RESPONSE} command. To avoid cross-phase compatibility problems, this service is only permitted to be used between a SIM and an ME that support the STK commands. The fact that an ME
supports specific STK commands is revealed when it sends the
\emph{TERMINAL PROFILE} command during SIM initialisation.

The SIM can make a variety of requests using the proactive
SIM service. Examples include: requesting the ME to display
SIM-provided text, initiating the establishment of on demand channels, and
providing local information from the ME to the SIM. The commands of interest here are \emph{GET CHANNEL STATUS}, which requests the ME to return the current status of all available data channel(s), and \emph{CLOSE CHANNEL}, which requests the ME to close the specified data channel. Both of these STK commands are marked as `class e', which means that an ME that supports `class e' STK commands is capable of executing both commands of interest \cite{TS_101_267}.
Although support of STK is optional for an ME, if an ME claims compliance with a specific GSM release then it is mandatory for the ME to support all functions of that release. Since 1998 almost all of the mobile phones produced have been STK enabled, and today every phone on the market supports STK \cite{STK}.

\section{RAND hijacking}  \label{hijacking}

We use the term \emph{RAND hijacking} to refer to the idea of
using the \emph{RAND}, sent from the network to the UE during
AKA, as a way of conveying information from the AuC to the
SIM\@. That is, instead of generating the \emph{RAND} at
random, it is generated to contain certain information; this
information is typically sent in encrypted form so that to an
eavesdropper it is indistinguishable from a random value.

This idea was apparently first described in a patent due
to Dupr\'{e} \cite{Dupre04}. However, the use Dupr\'{e} makes
of the idea is rather different to that proposed here. Later, Vodafone introduced the concept of a `special RAND'~\cite{S3-030463-Special_RAND} in 3GPP TSG document S3-030463. As for Dupr\'{e}, the purpose of the `special RAND' was completely different to that proposed here. The other published references to the notion appear in papers \cite{LTE-Identity-Privacy_Choudhury_2012,Khan15,Broek15} that independently propose
the use of RAND hijacking for improving the privacy properties
of GSM, 3G and 4G networks. As far as the authors are aware, no
previous authors have proposed the use of this technique for
providing mutual authentication in GSM networks.

\section{Server-to-SIM authentication}  \label{auth}

We now propose a way of using RAND hijacking to enable
authentication of the network to the SIM\@. For this to operate
the SIM must be programmed to support the scheme, as well as
possess certain (modest) additional data, as detailed below.
The AuC of the network issuing the `special' SIM must also
store certain additional data items for each such SIM, and must
generate its \emph{RAND} values in a special way for such SIMs.
No other changes to existing systems are required.

\subsection{Prerequisites}

In addition to sharing $K_i$, A3 and A8 (as required for
executing the standard GSM AKA protocol), the SIM and AuC must
both be equipped with the following information and functions:
\begin{itemize}
\item functions $f1$ and $f5$, where $f1$ is a MAC function
    and $f5$ is a cipher mask generation function, both
    capable of generating a 64-bit output;
\item a secret key $K_a$ to be used with functions $f1$ and
    $f5$ which should be distinct from $K_i$ --- to
    minimise memory requirements, $K_a$ and $K_i$ could,
    for example, both be derived from a single SIM-specific
    master key;
\item a 48-bit counter to be used to generate and verify
    sequence numbers\footnote{As in 3G, an AuC might choose
    to manage a single counter shared by all user accounts
    (see, for example, \cite{K356}).}.
\end{itemize}
The functions could be precisely the same as their counterparts
used in 3G (UMTS)\@.  Indeed, the function names and string
lengths have deliberately been made identical to those used in
3G systems to make implementation and migration as simple as
possible.

\subsection{Protocol operation}

The novel AKA protocol only differs from the `standard' GSM AKA
protocol (as described in section~\ref{AKA} above) in steps 3
and 6.  Thus, since these steps only involve the AuC and SIM,
it should be clear that the scheme is inherently  transparent
to the serving network and the ME\@.  We describe below how
these steps are changed.

\subsection{Revised steps}  \label{newsteps}

\begin{figure}
\centering
\includegraphics[height= 60mm, width=110mm]{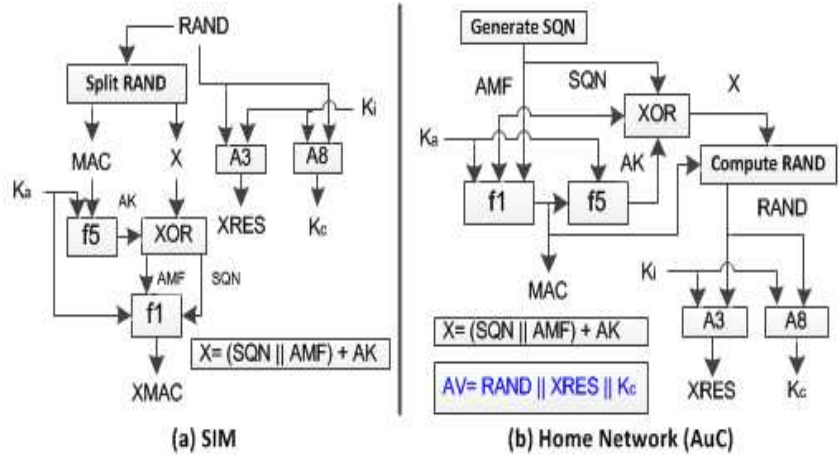}
\caption{Modifications at the SIM and the AuC}
\label{Modified_SIM-HN}
\end{figure}

Step 3 is changed to the following step 3*.  To generate a new
authentication triple, the AuC proceeds as follows (see Figure~\ref{Modified_SIM-HN}(b)).
\begin{enumerate}

\item[3.1] The AuC uses its counter value to generate a
    48-bit sequence number \emph{SQN}, which must be
    greater than any previously generated value for this
    user account.
\item[3.2] A 16-bit value \emph{AMF} is also generated,
    which could be set to all zeros, or could be used for
    purposes analogous to the \emph{AMF} value for 3G
    networks.
\item[3.3] A 64-bit tag value \emph{MAC} is generated using
    function $f1$, where
    \[ \mbox{\it MAC} = f1_{K_a}(\mbox{\it AMF}||\mbox{\it SQN}), \]
    and, as throughout, $||$ denotes concatenation of data
    items.
\item[3.4] A 64-bit encrypting mask \emph{AK} is generated
    using function $f5$, where
    \[ \mbox{\it AK} = f5_{K_a}(\mbox{\it MAC}). \]
\item[3.5] The 128-bit \emph{RAND} is computed as
    \[ \mbox{\it RAND} = ((\mbox{\it AMF}||\mbox{\it SQN})\oplus \mbox{\it
    AK}) || \mbox{\it MAC}, \] where, as throughout,
    $\oplus$ denotes the bitwise exclusive or operation.
\item[3.6] The \emph{XRES} and $K_c$ values are computed in
    the standard way, that is \emph{XRES} =
    A3$_{K_i}$(\emph{RAND}) and $K_c$ =
    A8$_{K_i}$(\emph{RAND}).
\end{enumerate}

Step 6 is changed to step 6*, as follows (see Figure~\ref{Modified_SIM-HN}(a)).

\begin{enumerate}
\item[6.1] On receipt of the 128-bit \emph{RAND} value, the
    SIM first splits it into two 64-bit strings $X$ and
    $\mbox{\it MAC*}$, where $X||\mbox{\it MAC*} =
    \mbox{\it RAND}$.
\item[6.2] A 64-bit decrypting mask $\mbox{\it AK*}$ is
    generated using function $f5$, where
    \[ \mbox{\it AK*} = f5_{K_a}(\mbox{\it MAC*}). \]
\item[6.3] A 16-bit string $\mbox{\it AMF*}$ and a 48-bit
    string $\mbox{\it SQN*}$ are computed as:
    \[ \mbox{\it AMF*} || \mbox{\it SQN*} = X \oplus \mbox{\it AK*}. \]
\item[6.4] A 64-bit tag \emph{XMAC} is computed as:
    \[ \mbox{\it XMAC} = f1_{K_a}(\mbox{\it AMF*}||\mbox{\it SQN*}). \]
\item[6.5] The recovered sequence number $\mbox{\it SQN*}$
    is compared with the SIM's stored counter value and
    \emph{XMAC} is compared with $\mbox{\it MAC*}$:
    \begin{itemize}
    \item if $\mbox{\it SQN*}$ is greater than the
        current counter value {\bf and} $\mbox{\it
        XMAC} = \mbox{\it MAC*}$, then:
        \begin{itemize}
        \item the network is deemed to be
            successfully authenticated;
        \item the SIM's counter value is updated to
            equal $\mbox{\it SQN*}$; and
        \item \emph{SRES} and $K_c$ are computed as
            specified in the current step 6;
        \end{itemize}
    \item if either of the above checks fail then:
        \begin{itemize}
        \item network authentication is deemed to
            have failed;
        \item the SIM's counter value is unchanged;
            and
        \item \emph{SRES} and $K_c$ are set to
            random values.
        \end{itemize}
    \end{itemize}
\end{enumerate}

It should be clear that, in step 6*, $\mbox{\it AK*}$,
$\mbox{\it AMF*}$, $\mbox{\it MAC*}$ and $\mbox{\it SQN*}$
should respectively equal the $\mbox{\it AK}$, $\mbox{\it
AMF}$, $\mbox{\it MAC}$ and $\mbox{\it SQN}$ values originally
computed by the AuC in step 3*.

\subsection{Design rationale}

The composition of the \emph{RAND} value in the above scheme
has been made as similar as possible to the 128-bit value
\emph{AUTN} used to provide server-to-UE authentication in the
3G AKA protocol. This is for two main reasons.  Firstly, as
stated above, by adopting this approach it is hoped that
implementation of, and migration to, this new scheme will be
made as simple as possible for network operators.  Secondly,
the 3G AKA protocol is widely trusted to provide
authentication, and it is hoped that trust in the novel scheme
will be maximised by adopting the same approach.

The only differences between the 3G \emph{AUTN} and the above
construction of \emph{RAND} are relatively minor, and are as
follows.
\begin{itemize}
\item In 3G, the \emph{AK} value is computed as a function
    of the the \emph{RAND}, whereas here it is necessarily
    only computed as a function of the last 64 bits of
    \emph{RAND}\@.  However, these last 64 bits are
    computed as a function of data which changes for every
    authentication triple, and hence the \emph{AK} should
    still do an effective job of concealing the content it
    is used to mask.
\item In 3G the \emph{AK} is only 48 bits long, and is only
    used to encrypt (mask) the \emph{SQN}\@.  Here we use
    it to mask the \emph{SQN} and the \emph{AMF}, to ensure
    that a `new style' \emph{RAND} is indistinguishable
    from an `old style' randomly generated \emph{RAND} to
    any party without the key $K_a$.
\item In 3G, the \emph{MAC} is computed as a function of
    the \emph{RAND}, \emph{SQN} and \emph{AMF}, whereas in
    the above scheme it is computed only as a function of
    \emph{SQN} and \emph{AMF}, again for obvious reasons.
    This is the only significant difference from the perspective of authenticating the network to a UE, but we
    argue below in section~\ref{security-analysis} that
    this change does not affect the security of the
    protocol.
\end{itemize}
The \emph{AUTN} checking process proposed here and that used in
3G are essentially the same.

One other issue that merits mention is the fact that it is
proposed that the SIM outputs random values if authentication
fails.  It is necessary for the SIM to output values of some
kind, since this is part of the existing SIM-ME protocol.  That
is, placeholder values are required.  It is important for
reasons discussed below that the SIM should \emph{not} output
the correct session key $K_c$. The only other `obvious'
placeholder values would be to use fixed strings, but the use
of random values seems less likely to be obvious if these
values are sent across the network (in the case of the
\emph{SRES} value) or used for encryption purposes (for $K_c$).
There may be advantages in not revealing to a casual
eavesdropper the fact that authentication has failed.

\section{Using the authentication results}  \label{using}

In the previous section we showed how the SIM can authenticate
the network; that is, as a result of step 6*, the SIM will know
whether or not the \emph{RAND} genuinely originates from the
AuC and is fresh. However, we did not describe any way for the
ME to know whether authentication has failed or succeeded
--- indeed, the ME will not understand the concept, as we
are assuming it is a `standard' GSM device.

\begin{figure}
\centering
\includegraphics[height= 65mm, width=55mm]{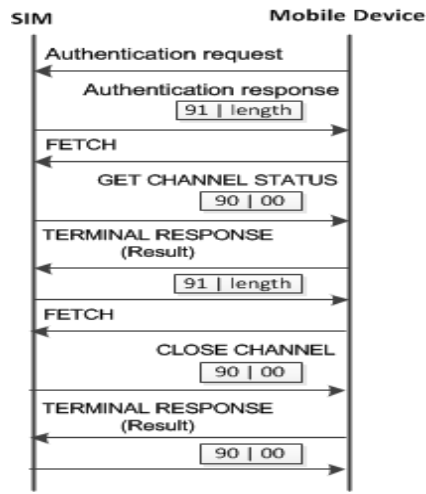}
\caption{SIM-ME interactions to drop the established connection}
\label{SIM-ME_Exchange}
\end{figure}

We propose that the proactive SIM feature described in
section~\ref{proactive} be used to achieve the desired
objective.  That is, in the event of a network authentication
failure, when sending the \emph{SRES} and
$K_c$ (in this case random) values back to the ME, the SIM should signal to the phone
that it has information to send. When, as a result, the ME sends the
\emph{FETCH} command to the SIM,
the SIM should respond with the \emph{GET CHANNEL STATUS} command to learn about the established channels in the present connection. Upon receiving the   
channel information in the \emph{TERMINAL RESPONSE} command, the SIM uses the response status byte in its response to request the ME to send a further \emph{FETCH} command. Once it receives the \emph{FETCH} command, the SIM responds with a \emph{CLOSE CHANNEL} command, specifying the channel information it received from the ME in response to its previous \emph{CHANNEL STATUS} command. The interactions between a SIM and an ME are summarised in Figure~\ref{SIM-ME_Exchange}. The STK commands issued by the SIM should cause the phone to drop the connection, and (hopefully)
prevent any attempted use of the \emph{SRES} or key $K_c$.

\section{Analysis}  \label{analysis}

\subsection{Deployment issues}

We next consider certain practical issues that may arise when
using the scheme proposed in section~\ref{newsteps}.

It seems that at least some GSM networks issue authentication
triples in batches (see section 3.3.1.1 of GSM 03.20 \cite{GSM_03_20}), thereby reducing the
inter-network communications overhead.  Currently, the order in
which GSM authentication triples are used does not matter.
However, under the scheme described above, triples must be used
in ascending order of \emph{SQN}\@.  This may seem problematic;
however, since the requirement to use authentication datasets
in the correct order already applies to the corresponding
5-tuples used in 3G, serving networks will almost certainly
already be equipped to do this.

In existing GSM networks it is possible, although restricted by the technical specifications \cite{GSM_03_20},
for serving networks to `re-use' authentication triples, i.e.\
to send the same \emph{RAND} value to a UE on multiple
occasions.  This will no longer work with the new scheme, since
the SIM will detect re-use of a \emph{RAND} value.  Arguably
this is good, since re-use of \emph{RAND} values is highly
insecure: such behaviour would allow the interceptor of a
\emph{RAND}/\emph{SRES} pair to impersonate a valid UE and
perhaps steal service at that UE's expense, an attack that
would be particularly effective in networks not enabling
encryption.

Finally note that, in order to fully implement the scheme as
described in section~\ref{auth}, MEs need to support `class e' STK commands, although, as discussed above, this proportion seems likely to be very high.  It is not clear what proportion of
mobile phones in current use support those STK commands.

\subsection{Security}  \label{security-analysis}

We divide our security discussion into three parts:
confidentiality and privacy issues, authentication of network
to SIM, and authentication of SIM to network.

\subsubsection{Confidentiality and privacy issues}

In `standard' GSM the \emph{RAND} value is randomly selected,
and so does not reveal anything about the identity of the phone
to which it is sent.  In the scheme proposed in
section~\ref{newsteps}, the \emph{RAND} is a function of a
SIM-specific key as well as a potentially SIM-specific
\emph{SQN} value.  However, the \emph{SQN} is sent encrypted,
and, assuming the functions $f1$ and $f5$ are well-designed, an
interceptor will not be able to distinguish an intercepted
\emph{RAND} computed according to the new scheme from a random
value.  Thus the scheme does not introduce a new threat to
identity confidentiality.

The new scheme does not change the way the data confidentiality
key $K_c$ is generated, so the strength of data confidentiality
is not affected.

\subsubsection{Network-to-SIM authentication}

The novel protocol for network-to-SIM authentication bears
strong similarities to the corresponding protocol for 3G\@. It
also conforms to the one-pass unilateral authentication
mechanism specified in clause 5.1.1 of 9798-4 \cite{K543,K658}.
All the protocols in this standard have been formally analysed
(and shown to be secure) by Basin, Cremers and Meier
\cite{Basin12}.

An interesting side observation deriving from the novel scheme
is that the 3G and 4G AKA protocols appear to be overly
complex. The randomly generated \emph{RAND} value sent from the
network to the SIM, which is used to authenticate the response
from the SIM to the network, is actually unnecessary, and the
\emph{AUTN} value could be used in exactly the same way as the
\emph{RAND} is currently.  Whilst such a change is not possible
in practice, it would have avoided the need for the AuC to
generate random values and saved the need to send $16$ bytes
in the AKA protocol.

It is interesting to speculate why this design redundancy is
present. It seems possible that the network-to-SIM
authentication was added as a completely separate protocol to
complement the GSM-type SIM-to-network authentication
mechanism, and no-one thought how the two mechanisms could be
combined and simplified (as in the mechanism we propose).

\subsubsection{SIM-to-network authentication}

The novel scheme does not affect how the existing
SIM-to-network authentication protocol operates, except that a
random \emph{RAND} is replaced by one which is a cryptographic
function of a sequence number.  The new-style \emph{RAND}
remains unpredictable to anyone not equipped with the key
$K_a$, and is deterministically guaranteed to be non-repeating
(a property that only holds in a probabilistic way for a random
\emph{RAND}). That is, it possesses precisely the qualities
required by the existing protocol, and hence the security of
SIM-to-network authentication is unaffected.

\subsection{Impact on known attacks}

We conclude our analysis of the protocol by considering how it
affects possible attacks on GSM networks.

\subsubsection{Fake network attacks}

As discussed in section~\ref{vulns}, if a phone joins a fake
GSM serving network, then this fake network can send any
\emph{RAND} value it likes as part of the AKA protocol, and the
UE will complete the process successfully.  If the network does
not enable encryption, then communications between the UE and
the network will work correctly, which could enable the network
to act as an eavesdropping man-in-the-middle by routing calls
from the captured UE via a genuine network. This will no longer
be true if the new scheme is implemented, since the SIM will
instruct the ME to drop the connection when supplied with a
non-genuine \emph{RAND} value.

Of course, it may be possible for a fake network to avoid the
AKA protocol altogether, and simply start communication with a
newly attached UE\@.  Whether MEs will accept unauthenticated
communication is currently not clear to the authors.

\subsubsection{Barkan-Biham-Keller attacks}

We next consider a particular type of fake network attack,
namely the Barkan-Biham-Keller attack outlined in
section~\ref{vulns}.  As described there, the attack requires
the re-sending of an `old' \emph{RAND} to a UE\@.  The new
scheme will clearly prevent such an attack, i.e.\ the
Barkan-Biham-Keller attack will be prevented, at least in most
practical scenarios.

\section{Relationship to the prior art}  \label{prior}

This is by no means the first practical proposal for enhancing GSM to
incorporate mutual authentication.  Indeed, the 3G AKA
protocol, discussed widely in this paper, can be regarded as
doing exactly that. Although several 3GPP TSG documents~\cite{S3-030542-AKA_Enhancement,S3-040534-UMTS_AKA_in_GSM} proposed the introduction of network authentication into the GSM network, none were adopted, presumably because of cost/feasibility issues. The Ericsson proposal~\cite{S3-040534-UMTS_AKA_in_GSM} suggested transferring authentication responsibility to the terminal by implementing the core of the UMTS AKA protocol entirely in software, which in turn raised other security threats. Other proposals have been made, including
by Kumar et al., \cite{Kumar06}.  However, all previous
proposals are completely impractical in that they would require
replacing all the GSM infrastructure.  Such a major change to
an existing very widely deployed scheme is simply not going to
happen. 

The most similar proposals to that given here are some of the
other schemes using RAND hijacking, summarised in
section~\ref{hijacking}.  In particular, van den Broek, Verdult
and de Ruiter \cite{Broek15} propose a similar structure for a
hijacked GSM RAND, in their case including a sequence number, a
new temporary identity for the SIM, and a MAC, all encrypted in
an unspecified way.  However, their objective is not to provide
authentication of the network to the SIM, but to provide a way
to reliably transport new identities from the AuC to the SIM.

\section{Concluding remarks}  \label{conclusion}

We have proposed a method for enhancing the GSM AKA protocol to
provide authentication of the network to the UE, complementing
the UE-to-network authentication already provided.  This
provides protection against some of the most serious threats to
the security of GSM networks.  This is achieved in a way which
leaves the existing serving network infrastructure unchanged,
and also does not require any changes to existing MEs (mobile
phones).  That is, unlike previous proposals of this general
type, it is practically realisable.

A number of practical questions remain to be answered, including the
proportion of MEs supporting `class e' STK commands, the behaviour of
MEs in networks which never perform the AKA protocol, and
whether serving networks can be relied upon to use GSM
authentication triples in the intended order. Discovering
answers to these questions remains as future work.

\bibliography{Crypto}

\end{document}